# Neutron scattering cross section of diamond nanoparticles

*Kenji* Mishima[1,2*], *Toshiya* Otomo[1,2], *Kazutaka* Ikeda[1,2], and *Hidetoshi* Ohshita[1,2]

[1]Neutron Science Section, Materials & Life Science Division, J-PARC Center, Tokai, Ibaraki, Japan
[2]Neutron Science Laboratory, High Energy Accelerator Research Organization, Tokai, Ibaraki, Japan

**Abstract.** Due to their large coherent scattering cross section, diamond nanoparticles (DNPs) are considered as a promising candidate material for a new neutron reflector. For investigation of scattering cross sections of packed samples, we have developed a technique for mechanical compression of DNP powder. Application of 220 MPa allowed us to increase the bulk density from 0.40 g/cm³ to 1.1 g/cm³. The differential cross sections of uncompressed and packed samples were measured using the high-intensity total diffractometer instrument NOVA at J-PARC, covering transfer wavenumbers ($q$) from 0.6 to 100 nm$^{-1}$. The $q$ dependence for the compressed sample agreed with the theoretical expectation derived from the Born approximation applied to homogeneous spheres with inclusion of a hard-sphere model to account for the inter-particle structure, whereas the results obtained from the powder sample disagreed. This implies that the theoretical description does not well represent the mesoscopic structure of the DNP powder sample.

## 1. Introduction

Owing to their unique properties, neutrons are widely employed as a probe for material sciences, imaging, fundamental physics and other applications. As an alternative to reactor based neutron sources for research, spallation neutron sources using megawatt-class proton accelerators have been developed or are under construction around the world [1-3]. However, the mechanical strength of the spallation target under proton bombardment has become a technical limitation towards higher-power neutron sources. Therefore, more effective neutron moderators and reflectors are required to increase available neutron fluxes.

Steyerl and Trüstedt observed that the inhomogeneous structure of electro-graphite leads to a strong coherent enhancement of the neutron scattering cross section. They also mentioned that this phenomenon lends itself for application as a neutron reflector for very slow neutrons [4]. Also diamond nanoparticles (DNPs) scatter neutrons strongly due to the bulk inhomogeneity induced by the particle form factor [5-9]. Recently, DNPs with a low abundance of chemical impurities and with narrow distributions of particle sizes have become inexpensive. Although large-angle diffusive scattering by DNPs is rather inefficient for neutron velocities larger than 100 m/s, fluxes of higher-energy neutrons from a moderator can possibly be increased due to diffusive scattering under small angles. For this, the bulk density of the diffusing material is an important parameter. DNPs are generally available as a powder with relatively low bulk density at the order of 0.40 g/cm³, as compared to the crystal density of 3.5 g/cm³ for pure diamond.

For the study presented in this paper, we developed a method to increase the bulk density to 1.1 g/cm³ by application of mechanical pressure using a steel die. As input for design work on neutron reflectors, we measured the differential cross sections of the powder and the compressed samples using the high-intensity total diffractometer instrument NOVA of the Material and Life Science Experimental Facility (MLF) at the Japan Proton Accelerator Research Complex (J-PARC) [10].

## 2. Single-particle scattering formalism

The coherent scattering cross section of a nanoparticle depends on the transfer wavenumber $q$, given by

$$q = 2k \sin\frac{\theta}{2}, \quad k = \frac{2\pi}{\lambda}, \qquad (1)$$

where $k$ and $\lambda$ are the neutron wavenumber and wavelength, respectively, and $\theta$ is the scattering angle. Coherent superposition of scattering amplitudes from individual nuclei leads to an enhanced cross section in directions of small $q$ (corresponding to smaller scattering angles for larger neutron energies). The amplitude of a neutron scattered by a uniform spherical particle can be calculated using the Born approximation and expressed as

$$\Phi(\theta) = -\frac{2m}{\hbar^2} U_0 R^3 \left( \frac{\sin(qR)}{(qR)^3} - \frac{\cos(qR)}{(qR)^2} \right), \qquad (2)$$

where $m$ is the neutron mass, $R$ and $U_0$ are the radius and the Fermi potential of the spherical particle, respectively, and $\hbar$ is Planck's constant [5]. The differential scattering cross section is given by

$$\frac{d\sigma}{d\Omega} = |\Phi(\theta)|^2. \qquad (3)$$

Figure 1 shows its $q$ dependence for spherical DNPs of three different sizes, as determined using Eqs. (1-3) for a DNP density of 3.1 g/cm³ taken from the manufacturer's specifications quoted in section 3.1. Enhancements of the differential cross section by more than 1,000 times with respect to a single C nucleus are visible in the low-$q$ region of less than 1 nm$^{-1}$.

---

[*] Corresponding author: kenji.mishima@kek.jp

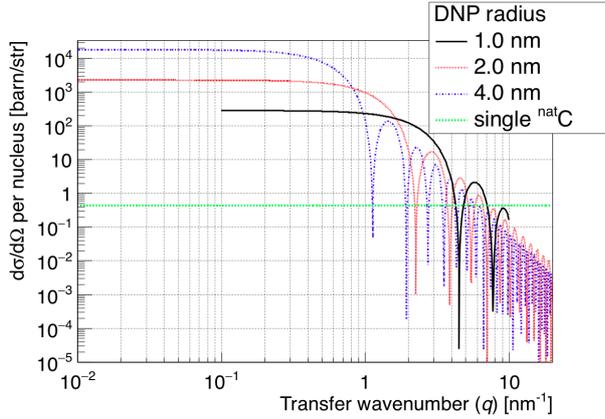

**Fig. 1.** Theoretical $q$ dependences of differential cross sections of DNPs for three particle radii indicated in the legend. The green line shows the scattering cross section of a single carbon nucleus (average for natural isotopic composition).

## 3. DNP samples

In this section, we describe the properties of the DNP base material and the preparation of the samples used for the differential cross section measurements.

### 3.1 DNP specifications

Base material of the samples for our study is the commercially available uDiamond® Molto Nuevo (Carbodeon), which comes with relatively high chemical purity [11]. The surface functional groups of the DNPs were removed to an extent that almost none of the carboxylic and amino groups are left over. The DNP specifications are listed in Table 1. The quoted density of 3.1 g/cm$^3$ is lower than the bulk density of diamond (3.5 g/cm$^3$) due to the graphitic structure of the DNP surface. Assuming a density of 2.2 g/cm$^3$ for the latter, the DNPs include 31% graphite.

**Table 1. DNP specifications**

| Parameter | Value |
|---|---|
| DNP name | uDiamond® Molto Nuevo |
| Radius | 2–3 nm |
| Purity | 97% |
| Water content | < 2wt% |
| Particle density | 3.1–3.2 g/cm$^3$ |
| Powder density | 0.4 g/cm$^3$ |
| Crystal lattice constant | 0.3573(5) nm |

Table 2 lists the impurities of the material as measured by X-ray fluorescence (XRF). Note that the values are quoted there for Molto instead of Molto Nuevo. However, since these two material grades differ only in the content of surface functional groups, the listed impurities can be expected to be the same. Their scattering can safely be neglected as they add only 0.02 barn to the scattering cross section per nucleus (corresponding to only 0.4% of a carbon nucleus, which is 5.551 barn [13]).

**Table 2. Contaminations in DNPs measured by XRF**

| Element | Concentration (weight ppm) |
|---|---|
| Na | 110 |
| Al | 220 |
| Si | 1200 |
| S | 3200 |
| Cl | 160 |
| Ti | 50 |
| Cr | 1900 |
| Fe | 380 |
| Br | < 50 |
| Hg | < 50 |
| Pb | < 50 |

The DNP size distribution is another important parameter. We deduced this information from a transmission electron microscope (TEM) image (see Fig. 2). The measured average DNP radius of 2.8 nm with a standard deviation of 1.1 nm is consistent with the manufacturer's specification given in Table 1.

### 3.2 DNP sample fabrication

#### 3.2.1 Powder DNP sample

The first sample for our neutron scattering study consisted of uncompressed DNP powder in an aluminum cell with a volume of $20 \times 20 \times 1.0$ mm$^3$. The cell was made of the Al050 aluminum alloy; the lid and the container were 0.5 mm thick each, so that the total thickness of Al intersected by the beam was 1 mm. The sample was prepared by filling the powder into the cell, scraping off the excess and sealing it with the lid. The mass difference between the full and the empty cell, 0.1616 g, corresponds to a DNP bulk density of $0.40 \pm 0.04$ g/cm$^3$, with the uncertainty being estimated based on the machining accuracy. Figure 3 shows a photograph of the container filled with the powder sample before closing it. The contribution of the aluminum to the scattering signal was evaluated from measurements using another empty sample container.

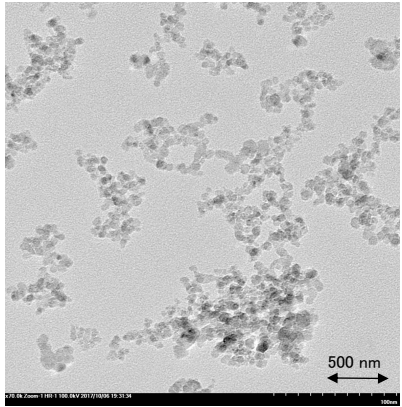

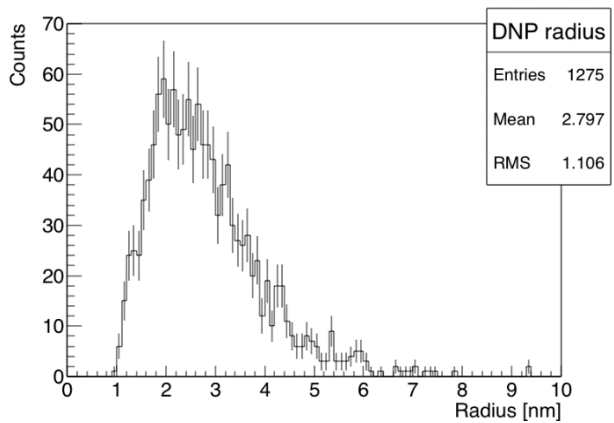

**Fig. 2.** TEM image of the DNP sample material (top). Particle radius distribution deduced from this TEM image (bottom).

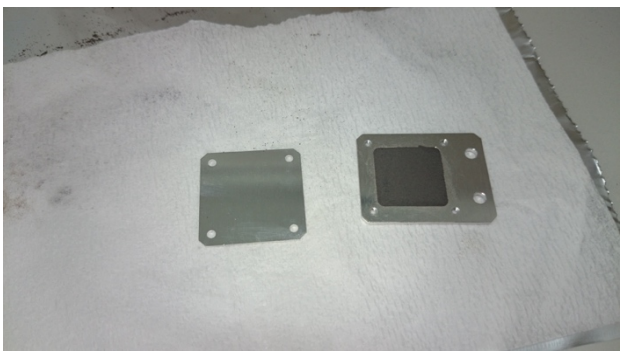

**Fig. 3.** Sample container filled with the DNP powder.

### 3.2.2 Compressed DNP sample

For use as a neutron reflector, the DNP bulk density should ideally be approximately half of the particle density. This requires a significant increase of the bulk density of the powder (delivered with only 0.4 g/cm$^3$). In order to prepare a packed sample for our neutron scattering study we compressed the powder to a solid slab, using the die shown in Fig. 4. This tool made of 10-mm thick steel sheets was composed of a bottom and four side parts. A sample of 0.148 g DNP powder was filled in the central square hole with an area of 10 mm × 10 mm and a depth of 10 mm. It was compressed by a convex lid, using the hydraulic press shown in Fig. 5. The applied pressure was 220 MPa, as calculated from the cylinder area (7.16 cm$^2$) and the area of the convex lid (10 mm$^2$). The measured thickness of the packed DNP sample, 1.3 ± 0.1 mm, corresponds to a bulk density of 1.14 ± 0.09 g/cm$^3$. The last step of the sample preparation and the final product are shown in Figs. 6 and 7.

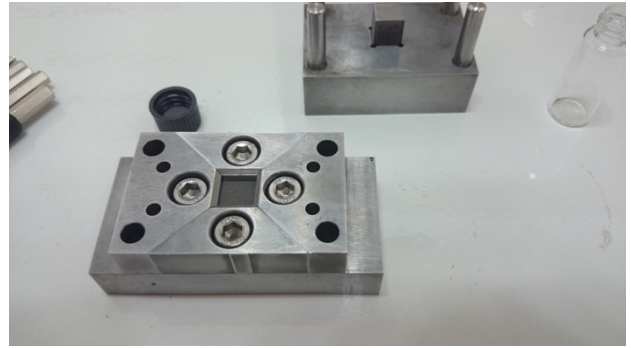

**Fig. 4.** Pressurizing die.

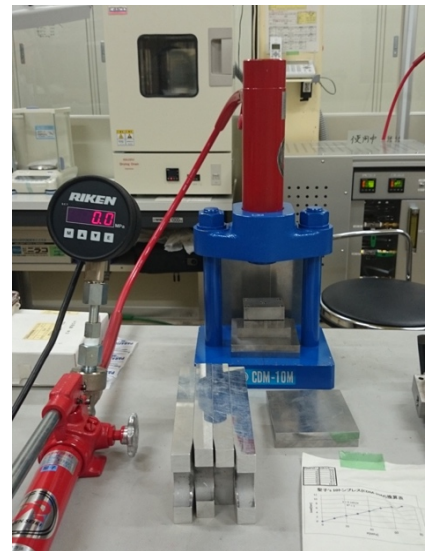

**Fig. 5.** Hydraulic press used for preparation of the packed DNP sample.

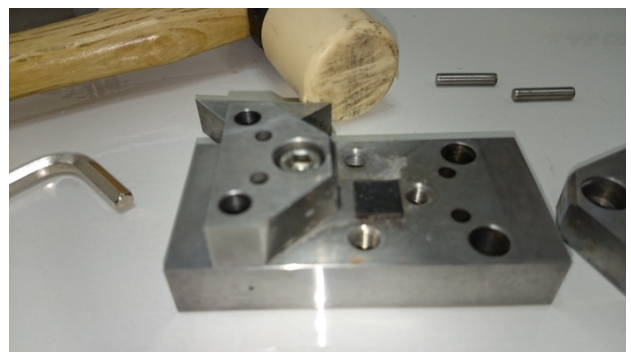

**Fig. 6.** Extraction of the packed DNP sample from the die.

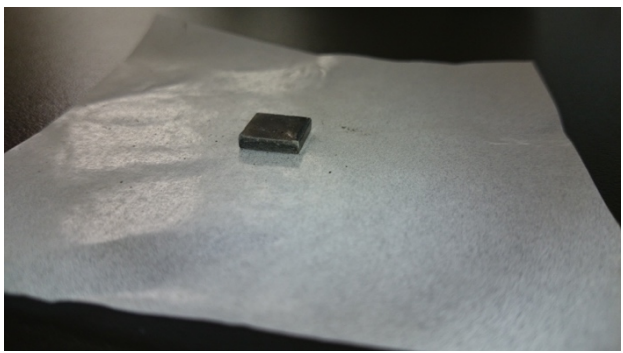

**Fig. 7.** Packed DNP sample.

## 4. Measurements

Measurements of differential neutron scattering cross sections were performed using the diffractometer NOVA at MLF at J-PARC [10]. A sample environment was prepared for automatized sample change. The instrument can measure neutron scattering over a wide $q$ range of $0.1 < q < 1000$ nm$^{-1}$ with 0.35% resolution [12]. NOVA has five neutron detector banks placed at different angles. Each bank consist of 800-mm long position-sensitive $^3$He proportional counters. In the reported experiments, only three banks were employed, covering the forward direction ($0.7^0 – 9^0$) and angular ranges about $20^0$ and $45^0$ (the $90^0$ and back-scattering banks of the instrument were not used). Two low-opacity gas electron multiplier (GEM) detectors were used as beam monitors [13], being positioned at the inlet and the outlet of the sample position. The MLF provides pulsed neutron beams with 25-Hz repetition rate. Using the time of flight (TOF) one can do wavelength-resolved measurements of scattered neutrons. To avoid frame overlap (the problem that, notably in small-angle scattering measurements, the signal may be contaminated by slow neutrons generated in the previous pulse), a "tail cutter" installed upstream was used to remove slow neutrons with velocities in the range 196–500 m/s. Figure 8 shows the spectrum measured with the inlet beam monitor. The kink at 0.8-nm wavelength (corresponding to 500 m/s) is due to the tail cutter.

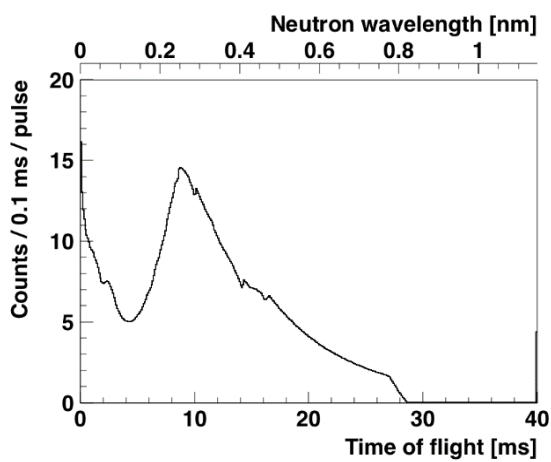

**Fig. 8.** Neutron TOF spectrum measured using the inlet beam monitor at NOVA.

Neutron scattering measurements were performed for the powder sample and the packed, solid sample described above, and for empty cells of each. Additional measurements with a vanadium plate served for calibration of the detectors which was needed to obtain the cross sections on an absolute scale. The scattering by V being almost entirely incoherent (5.1 barn [14] but only 0.0184 barn coherent cross section) it is isotropic with only a small distortion due to inelastic scattering of 8% at room temperature (293 K) [15]. The latter is taken into account in the data analysis as a systematic uncertainty. The V plate had an area of $37 \times 30$ mm$^2$ and a thickness of 1.5 mm. To correctly compare the samples of different sizes, Cd collimators were placed close in front of the sample position, preparing a beam of 10 mm × 10 mm at the sample. The measurements were performed for $3 \times 10^3$ s ($7.0 \times 10^3$ kicker pulses) for each DNP sample and for each empty cell, and for $2 \times 10^4$ s ($4.4 \times 10^5$ kicker pulses) for the V calibration plate. The beam power of the proton accelerator was 150 kW. The number of protons per pulse was stable within 0.1% throughout all measurements.

## 5. Analysis and results

The detected neutrons were classified in terms of their $q$ values, which were determined from the detected position in the $^3$He detectors and the measured TOF. The latter was also used to restrict the velocity range of neutrons to be further analyzed to 500 – 6000 m/s. Faster neutrons were not taken into account because they would not be analyzed correctly due to a high-pass cut-off of Cd (~10,000 m/s) from where on the Cd collimations lose their opacity. The backgrounds were subtracted using measurements with the blank cells normalized to the number of kicker pulses. Absolute normalization of the cross sections included, besides the detector calibration using the V plate, normalization to the mass of the material exposed to the neutron beam. The result had 4% uncertainty based on the machining accuracy of the Cd collimators.

The obtained $q$ dependence of the differential cross section for the compressed sample is shown in Fig. 9. The data points are indicated by black points in the upper graph. The error bars show statistical errors, which are dominated by the data taken with the V plate. The red line is a data smoothing curve. The difference of data points from the curve is shown in the lower part of Fig. 9. The data for $q < 10$ nm$^{-1}$ show a systematic separation of points which we believe to be due to imperfect determination of the $q$ region from the TOF and the position of neutron detection. In particular, the accuracy of the position of detection becomes worse at the edges of the detector banks. Almost all data points agree with the smoothing curve to within 20%. Since the vanadium calibration does not correct this effect, we allocated the discrepancy of 20% as a systematic uncertainty. Including other uncertainties, the total uncertainty of the absolute scattering cross section was budgeted as 22%.

The black dashed line represents the differential scattering cross section of carbon with natural isotopic abundance (0.442 barn/str). The data points in the $q$ region larger than 20 nm$^{-1}$, where the scattering due to the nanoparticle structure becomes negligible, agree within 5% with the isotropic scattering as expected from individual carbon nuclei. This result shows that in good approximation the sample consists of pure carbon. A contamination of the sample by 2 wt% of $H_2O$, which is the maximum specification, would increase the scattering cross section by 38%. Therefore, taking into account the systematic uncertainty of the measurement of 22%, the $H_2O$ contamination in the measured sample can be estimated as being less than 1%.

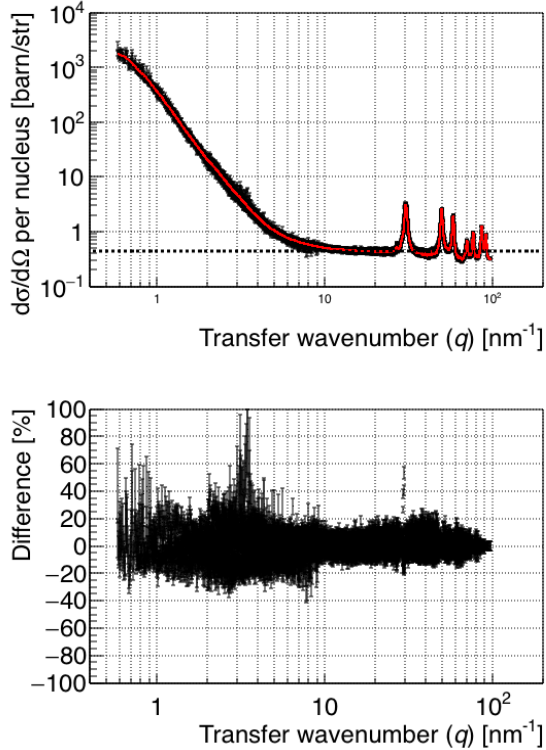

Fig. 9. (Top) NOVA data points for differential cross section of compressed DNP sample. The red line is a data smoothing curve. The horizontal black dashed line represents the differential scattering cross section of the carbon nucleus at natural isotopic abundance. (Bottom) Difference from the smoothing curve.

In top of Fig. 10, the smoothed curves of the compressed sample and of the powder sample, which was analyzed in the same manner, are plotted together. Peaks occurring in the range 30 to 60 nm$^{-1}$ are, in ascending order, due to Bragg diffraction from the diamond crystal planes (111), (220), and (311). Sharper spikes apparent in the high-$q$ region for the powder sample are due to inaccurate subtraction of the Bragg reflections by the Al container. In the low-$q$ region, there is no significant difference (<20%) between the results for the powder and the solid samples.

The cross section calculated using Eqs. (1-3) and taking into account the measured particle size distribution (see Fig. 2) is shown in Fig. 10 as a dash-dotted green line. The deviations from the experimental results observed for $q < 1$ nm$^{-1}$ are due to inter-particle correlations between the nanoparticles. To take the latter into account, we included the inter-particle structure factor $S'(q)$ of the Percus-Yevick (PY) hard sphere model [16]:

$$S'(q) = 1 + \frac{|\sum_i n_i \Phi(q, R_i)|^2}{\sum_i |n_i \Phi(q, R_i)|^2} \left[ \frac{n_i S(q, R_i, \eta)}{\sum_i n_i} - 1 \right], (4)$$

This correction function [17] takes into account the DNP size distribution shown in Fig. 2, with $n_i$ denoting the weight factor of particle radius $R_i$. The factor $S(q, R, \eta)$ approximates the interference factor for the PY hard sphere model [18]. The volume fraction $\eta$ is obtained from the bulk density of the sample divided by the DNP particle density of 3.1 g/cm$^3$. For the compressed and the powder samples it thus had the values 0.35 and 0.13, respectively. After application of the inter-particle correction, the compressed sample showed good agreement (<20%) down to 0.6 nm$^{-1}$. However, a significant difference (80%) remained for the powder sample. We believe this to be due a higher order (mesoscopic) structure, such as aggregation of the DNPs.

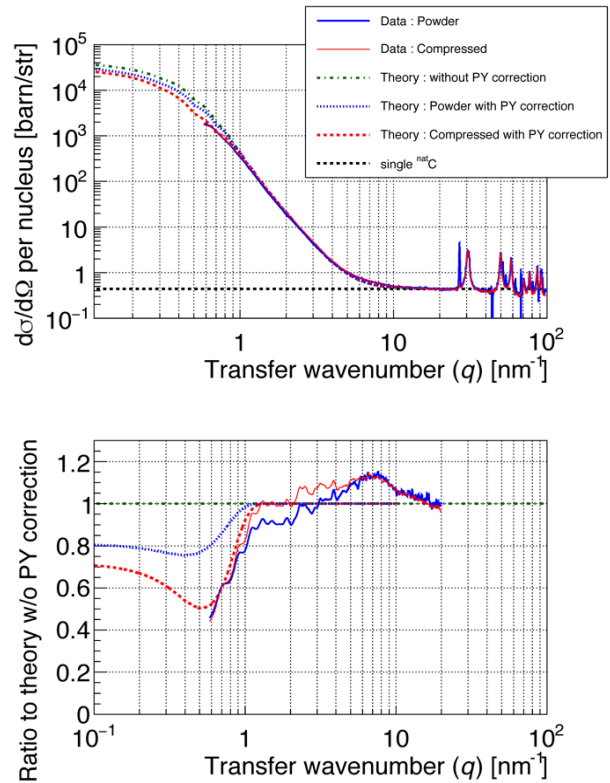

Fig. 10. (Top) Smoothed differential scattering cross sections for the compressed (solid red) and powder (solid blue) samples. The three theoretical curves display the cross sections with PY hard-sphere correction for the compressed (dot-blue) and the powder (dash-red) sample, and without correction (dot-dash green). The black dashed line shows the cross section of a single carbon nucleus. (Bottom) Ratios of theoretical cross sections with and without PY correction, together with the smoothed experimental data (line styles corresponding to the top figure).

## 6. Summary

DNPs can be expected to become employed for a new neutron reflector because of their large coherent scattering cross section. For this application a higher bulk density of the DNPs than in the commercially available powders is advantageous. We therefore developed a method to apply a pressure of 220 MPa with a hydraulic pump on a steel die, by which we succeeded to increase the bulk density from 0.4 to 1.1 g/cm$^3$.

The absolute differential scattering cross sections of a powder and a packed DNP sample were measured using the instrument NOVA at J-PARC, covering the $q$ range 0.6-100 nm$^{-1}$. Using a calibration of the detectors with vanadium a total accuracy of 22% was achieved. Results for the cross section per nucleus at high $q$ values around 20 nm$^{-1}$, where the particle structure has no influence, agreed within 20% with that of a single carbon nucleus, i.e., 0.442 barn/str. This shows that the $H_2O$ contamination of the DNP samples is less than 1wt%. In the $q$ range 1-10 nm$^{-1}$ the differential cross sections of the DNP samples, powder and packed, were consistent with the model of homogeneous spherical particles with size distribution as obtained by a TEM image. The analysis included a correction due to inter-particle correlations which was calculated using the hard-sphere model applied to the samples with different densities. For $q < 1$ nm$^{-1}$, this model agrees well with the experimental results for the compressed sample. The disagreement for the powder we believe to be caused by a mesoscopic structure, such as aggregation of the DNPs.

## Acknowledgements

We are grateful to members of the MLF Research Project, "Neutron source development for high intense cold neutron," for discussion of DNP treatments. We also thank T. Hattori, A. Sano, and S. Kawamura for assistance with DNP compression, and N. L. Yamada and H. Endo for fruitful discussions of the small angle scattering technique. This research was supported by a JSPS Grant-in-Aid for Exploratory Research (15K13413), for Scientific Research(S) (18H05230) and or (A) 18H03702. The experiment was approved by the Neutron Science Proposal Review Committee of J-PARC/MLF (Proposal Nos. 2016A0283 and 2016B0187).